\documentclass[english,conference]{IEEEtran}
\usepackage[T1]{fontenc}
\usepackage[latin9]{inputenc}
\usepackage{amsthm}
\usepackage{amsmath}
\usepackage{amssymb}
\usepackage{graphicx}

\makeatletter
  \theoremstyle{definition}
  \newtheorem{defn}{\protect\definitionname}
  \theoremstyle{plain}
  \newtheorem{prop}{\protect\propositionname}
\theoremstyle{plain}
\newtheorem{thm}{\protect\theoremname}

\usepackage{babel}
\usepackage{balance}
\usepackage{cite}

\makeatother

\usepackage{babel}
  \providecommand{\definitionname}{Definition}
  \providecommand{\propositionname}{Proposition}
\providecommand{\theoremname}{Theorem}

\begin{document}

\title{Competition vs. Cooperation: A Game-Theoretic Decision Analysis for
MIMO HetNets}

\author{\IEEEauthorblockN{Mathew Goonewardena \IEEEauthorrefmark{1}, Xin Jin \IEEEauthorrefmark{3}, Wessam Ajib \IEEEauthorrefmark{2}, and Halima Elbiaze\IEEEauthorrefmark{2}}    \IEEEauthorblockA{\IEEEauthorrefmark{1}\IEEEauthorrefmark{2} Department of Computer Science, Université du Québec à Montréal (UQAM), Canada,  } 
\IEEEauthorblockA{\IEEEauthorrefmark{3}  Department of Electrical and Computer Engineering, Université Pierre-et-Marie-Curie (UPMC), France,  } 
{\IEEEauthorrefmark{1}  mathew-pradeep.goonewardena.1@ens.etsmtl.ca}, {\IEEEauthorrefmark{3} xin.jin@etu.upmc.fr}, {\IEEEauthorrefmark{2}\{ajib.wessam, elbiaze.halima\}@uqam.ca}}
\maketitle
\begin{abstract}
This paper addresses the problem of competition vs. cooperation in
the downlink, between base stations (BSs), of a multiple input multiple
output (MIMO) interference, heterogeneous wireless network (HetNet).
This research presents a scenario where a macrocell base station (MBS)
and a cochannel femtocell base station (FBS) each simultaneously serving
their own user equipment (UE), has to choose to act as individual
systems or to cooperate in coordinated multipoint transmission (CoMP).
The paper employes both the theories of non-cooperative and cooperative
games in a unified procedure to analyze the decision making process.
The BSs of the competing system are assumed to operate at the\emph{
}maximum expected sum rate\emph{ }(MESR)\emph{ }correlated equilibrium\emph{
}(CE), which is compared against the value of CoMP to establish the
stability of the coalition. It is proven that there exists a threshold
geographical separation, $d_{\text{th}}$, between the macrocell user
equipment (MUE) and FBS, under which the region of coordination is
non-empty. Theoretical results are verified through simulations.
\end{abstract}

\section{Introduction}

Small cells are an easily deployable solution to the increasing demand
for capacity. Underlay small cells improve the capacity of the network
through frequency reuse and higher link gains due to shorter distances
to the user equipment (UE). On the downside the unplanned deployment
of small cells in the larger cell structure creates unforeseen interference
conditions. Such dynamic interference situations require novel solutions
\cite{Zahir2013}.

\emph{Coordinated multipoint transmission} (CoMP) introduces dynamic
interaction between multiple cells to increase network performance
and reduce interference. In our research we consider the CoMP scheme
of joint transmission (JT) \cite{Sawahashi2010}. We begin with the
hypotheses that JT must be a rational decision, which is profitable
for both macro- and femto-systems, since these systems may belong
to independent operators/users. In human interactions, cooperation
among a group is justifiable if all the members are better off in
that group than if they were in any other group structure among themselves.
This rational behavior is embedded in the solution concept of \emph{core}
in coalition formation games\emph{. }

Past research of heterogeneous networks (HetNets) of macro- femtocells,
has used both non-cooperative and cooperative games. In \cite{Kang2012}
a Stackelberg game is formulated where pricing is employed to move
the equilibria towards a tolerable interference level for the \emph{macrocell
base station} (MBS). In \cite{Buzzi2012} a potential game based analysis
of \emph{Nash equilibrium} (NE) of power and subcarrier allocation,
for a multicell interference environment is presented. In \cite{Huang2011}
power distribution over resource blocks of cognitive \emph{femtocell
base stations} (FBSs) is analyzed for their \emph{correlated equilibrium}
(CE). In \cite{Bennis2012,Lai2013} $\epsilon$-correlated equilibrium
solution is presented for underlayed femtocells to minimize interference
to the macro-system. CE is the form of equilibrium used in this paper
as well.

In \cite{Pantisano2011,Zhang2013} coalition formation games with
externalities are used to group the femtocells to mitigate collisions
and reduce interference. In \cite{Pantisano2012} a coalition game
together with the solution concept of \emph{recursive core }is used
to model the cooperative interaction between \emph{macrocell user
equipment} (MUE) and \emph{femtocell user equipment} (FUE). They conclude
that forming of disjoint coalitions increases the rates of both MUE
and FUE. In \cite{Ma2013} a coalition formation game is employed
to partition a dense network of femtocells to minimize interference
where they introduce a polynomial time algorithm for group formation.
In \cite{Mathur2008} both \emph{transferable utility} (TU) and \emph{non-transferable
utility} (NTU) coalition formation games are used for cooperation
of receivers and transmitters in an interference environment.

This paper is set apart from the above related research, since it
brings together both theories of non-cooperative and coalition formation
games to model femto-maro interaction in CoMP. A similar analysis
but, for non-CoMP case, is presented in \cite{Larsson2008}. The terms
non-cooperative and cooperative are in accordance to their use in
the game theory literature whereas the terms coordination, CoMP, and
JT are used synonymously.

The rest of the paper consists of the system model in Section \ref{sec:System-Model},
game-theoretic formulation and solution in Section \ref{sec:Core-Solution},
simulation results in Section \ref{sec:Numerical-Results} and conclusion
with summary in Section \ref{sec:Conclusion}.

\section{System Model\label{sec:System-Model}}

\begin{figure}
\includegraphics[scale=0.11]{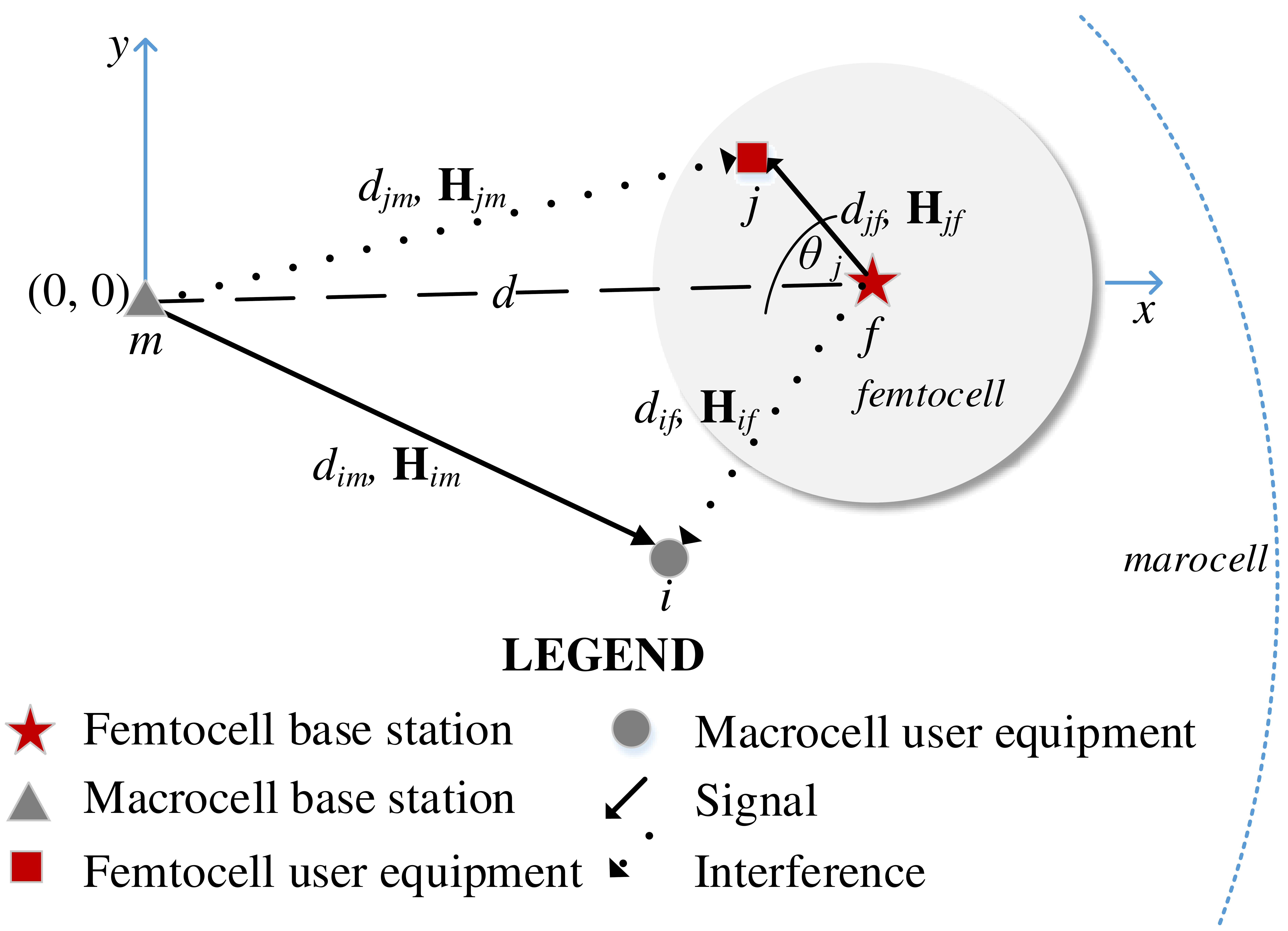}\caption{System model. The attributes on arrows indicate the effective distance
between elements and their respective channel matrices.\label{fig:System-model}}
\end{figure}

The paper considers the downlink transmission of a two tier HetNet,
which consists of a single MBS $m$ and a single FBS $f$, separated
by a distance $d>0$. Each \emph{base station }(BS) has an active
\emph{user equipment }(UE). It is possible that the BSs serve more
than one user but the assumption is that at any given instant each
BS transmits to only one selected user. The two BSs each possesses
$T$ number of transmit antennas while each UE possesses $R$ number
of receive antennas. Fig. \ref{fig:System-model} depicts the system
model. The origin of the plane is at MBS. We define two modes of operation,
namely uncoordinated and coordinated. In uncoordinated mode the two
BSs act as separate transmitters where MBS serves MUE while FBS serves
FUE. On the contrary if the two cells conform to the coordinated mode,
then the two BSs cooperate through CoMP.

The channel model includes large scale signal attenuation as a function
of distance. Channel gain matrix is multiplied by a magnitude, which
is a path loss function of distance between BS and UE \cite{Zhang2010}.
The received baseband equivalent signal $\mathbf{y}_{i}$ (resp. $\mathbf{y}_{j}$)
at MUE $i$ (resp. FUE $j$) for uncoordinated transmission over Gaussian
channel are 
\begin{eqnarray}
\mathbf{y}_{i} & \triangleq & d_{im}^{-\alpha}\mathbf{H}_{im}\mathbf{V}_{m}\mathbf{s}_{i}+d_{if}^{-\alpha}\mathbf{H}_{if}\mathbf{V}_{f}\mathbf{s}_{j}+\mathbf{n}_{i},\label{eq:MUErxsig}\\
\mathbf{y}_{j} & \triangleq & d_{jf}^{-\alpha}\mathbf{H}_{jf}\mathbf{V}_{f}\mathbf{s}_{j}+d_{jm}^{-\alpha}\mathbf{H}_{jm}\mathbf{V}_{m}\mathbf{s}_{i}+\mathbf{n}_{j},\label{eq:FURrxsig}
\end{eqnarray}
where $d_{im},\, d_{if},\, d_{jf},\, d_{jm}\geq0$ are the effective
distances between the respective indexed elements, $\mathbf{H}_{im}$
is the $R\times T$ complex valued channel gain matrix from MBS $m$
to MUE $i$ and $\mathbf{H}_{if},$ $\mathbf{H}_{jf},$ $\mathbf{H}_{jm}$
are interpreted analogously. The matrix $\mathbf{V}_{m}$ (resp. $\mathbf{V}_{f}$)
is the precoder at MBS (resp. FBS). The independent symbol vector
of unit variance at MBS $m$ (resp. at FBS $m$) to MUE $i$ (resp.
to FUE $j$) is denoted by $\mathbf{s}_{i}$ (resp. $\mathbf{s}_{j}$).
The exponent $\left(-\alpha\right)$, where $\alpha$ is a positive
real valued scalar, accounts for path loss, and $\mathbf{n}_{i},$
$\mathbf{n}_{j}$ are circular symmetric, uncorrelated additive withe
Gaussian noise (AWGN) vectors.

The achievable rate, treating interference as noise, of the macro
system, $R_{\text{uc}}^{m}$, is given by (\ref{eq:MUEcap}) \cite{Christensen2008}.
\begin{equation}
R_{\text{uc}}^{m}\triangleq\log\det\left(\mathbf{I}_{R}+\frac{d_{im}^{-2\alpha}\mathbf{H}_{im}\mathbf{V}_{m}\mathbf{V}_{m}^{H}\mathbf{H}_{im}^{H}}{d_{if}^{-2\alpha}\mathbf{H}_{if}\mathbf{V}_{f}\mathbf{V}_{f}^{H}\mathbf{H}_{if}^{H}+\sigma^{2}\mathbf{I}_{R}}\right),\label{eq:MUEcap}
\end{equation}

\begin{equation}
R_{\text{uc}}^{f}\triangleq\log\det\left(\mathbf{I}_{R}+\frac{d_{jf}^{-2\alpha}\mathbf{H}_{jf}\mathbf{V}_{f}\mathbf{V}_{f}^{H}\mathbf{H}_{jf}^{H}}{d_{jm}^{-2\alpha}\mathbf{H}_{jm}\mathbf{V}_{m}\mathbf{V}_{m}^{H}\mathbf{H}_{jm}^{H}+\sigma^{2}\mathbf{I}_{R}}\right).\label{eq:FUEcap}
\end{equation}
Above $\sigma^{2}$ is variance of circular symmetric noise and $\mathbf{I}_{R}$
is the $R\times R$ identity matrix. For a matrix $\mathbf{X}$ in
the complex field, $\mathbf{X}^{H}$ denotes the Hermitian transpose.
Analogously we define the achievable rate, $R_{\text{uc}}^{f}$, of
the femto-system (\ref{eq:FUEcap}).

Now suppose that the two BSs coordinate through JT. The coordination
is such that, FBS must transmit to both UE their respective symbols.
It is possible to extend this model to include the case where both
BSs transmit to both UE. The paper only consider MUE receiving JT
since FUE are mostly home/office users who are less mobile and they
have higher downlink gains whereas MUE may be highly mobile and operate
under high signal fading and interference. The received signals at
MUE and FUE in coordinated transmission are then given by (\ref{eq:coopMUEsig})
and (\ref{eq:coopFUEsig}) respectively. Note matrix augmentation
in (\ref{eq:coopMUEsig}).

\begin{alignat}{1}
\mathbf{y}_{i} & \triangleq d_{im}^{-\alpha}\mathbf{H}_{im}\mathbf{V}_{m}\mathbf{s}_{im}+d_{if}^{-\alpha}\mathbf{H}_{if}\mathbf{V}_{if}\mathbf{s}_{if}+\mathbf{n}_{i},\label{eq:coopMUEsig}\\
\mathbf{y}_{j} & \triangleq d_{jf}^{-\alpha}\mathbf{H}_{jf}\mathbf{V}_{jf}\mathbf{s}_{j}+d_{jm}^{-\alpha}\mathbf{H}_{jm}\mathbf{V}_{m}\mathbf{s}_{im}+\mathbf{n}_{j},\label{eq:coopFUEsig}\\
R_{\text{c}}^{f} & \triangleq\log\det\left(\mathbf{I}_{R}+\frac{d_{jf}^{-2\alpha}\mathbf{H}_{jf}\mathbf{V}_{jf}\mathbf{V}_{jf}^{H}\mathbf{H}_{jf}^{H}}{d_{jm}^{-2\alpha}\mathbf{H}_{jm}\mathbf{V}_{m}\mathbf{V}_{m}^{H}\mathbf{H}_{jm}^{H}+\sigma^{2}\mathbf{I}_{R}}\right).\label{eq:FUEcpcoop}
\end{alignat}

Above $\mathbf{V}_{if}$ (resp. $\mathbf{V}_{jf}$) is the precoder
matrix at FBS for MUE (resp. FUE), $\mathbf{V}_{m}$ is the MBS precoder.
The signal model assumes that the precoders $\mathbf{V}_{if}$ and
$\mathbf{V}_{jf}$ are such that there is no interuser interference
from FBS to the two UE. To that end block diagonalization (BD) can
be employed at FBS \cite{Hadisusanto2008}. The independent unit variance
transmission streams from MBS and FBS to MUE are $\mathbf{s}_{im}$
and $\mathbf{s}_{if}$. Then the achievable rates of FUE $R_{\text{c}}^{f}$,
and MUE $R_{\text{c}}^{m}$, for the coordinated transmission scheme
are given by (\ref{eq:FUEcpcoop}) and (\ref{eq:MUEcapcoop}) respectively
\cite{Tse2005}. This paper consider that the precoders at the two
BSs are chosen from a finite predefined code-book. The finite code
book model not only affords a finite action space game, but also reflects
the systems in practical implementations such as LTE, which define
a finite code-book. 

\begin{figure*}
\begin{equation}
R_{\text{c}}^{m}\triangleq\log\det\left(\mathbf{I}_{R}+\frac{d_{im}^{-2\alpha}}{\sigma^{2}}\mathbf{H}_{im}\mathbf{V}_{m}\mathbf{V}_{m}^{H}\mathbf{H}_{im}^{H}\right.+\left.\frac{d_{if}^{-2\alpha}}{\sigma^{2}}\mathbf{H}_{if}\mathbf{V}_{if}\mathbf{V}_{if}^{H}\mathbf{H}_{if}^{H}\right),\label{eq:MUEcapcoop}
\end{equation}

\begin{equation}
0\prec\mathbf{I}_{R}+d_{im}^{\prime-2\alpha}\left(d_{if}^{-2\alpha}\mathbf{B}+\mathbf{C}\right)^{\frac{-1}{2}}\mathbf{A}\left(d_{if}^{-2\alpha}\mathbf{B}+\mathbf{C}\right)^{\frac{-1}{2}}\preceq\mathbf{I}_{R}+d_{im}^{-2\alpha}\left(d_{if}^{-2\alpha}\mathbf{B}+\mathbf{C}\right)^{\frac{-1}{2}}\mathbf{A}\left(d_{if}^{-2\alpha}\mathbf{B}+\mathbf{C}\right)^{\frac{-1}{2}},\label{eq:claim2ineq}
\end{equation}

\begin{equation}
0\prec\mathbf{I}_{R}+d_{im}^{-2\alpha}\mathbf{A}^{\frac{1}{2}}\left(d_{if}^{-2\alpha}\mathbf{B}+\mathbf{C}\right)^{-1}\mathbf{A}^{\frac{1}{2}}\preceq\mathbf{I}_{R}+d_{im}^{-2\alpha}\mathbf{A}^{\frac{1}{2}}\left(d_{if}^{\prime-2\alpha}\mathbf{B}+\mathbf{C}\right)^{-1}\mathbf{A}^{\frac{1}{2}}.\label{eq:claim2ineq2}
\end{equation}
\end{figure*}

\section{Core Solution\label{sec:Core-Solution}}

Now the paper presents two non-cooperative games one for the uncoordinated
system, $\mathtt{G}_{1}$ and one for the coordinated system, $\mathtt{G}_{2}$.
The relation between $\mathtt{G}_{1}$ and $\mathtt{G}_{2}$ is established
in the ensuing development. Both games have identical set of players
$\mathcal{\mathcal{N}\triangleq\left\{ \text{MBS, FBS}\right\} }$,
i.e., the two BSs. The action spaces of the players are their precoder
code-books. In the uncoordinated case (resp. coordinated case) the
sets of precoders of MBS and FBS are denoted by $\mathcal{A}_{m\text{uc}}$
and $\mathcal{A}_{f\text{uc}}$ (resp. $\mathcal{A}_{m\text{c}}$
and $\mathcal{A}_{f\text{c}}$) respectively. Let $\mathbf{0}\notin\mathcal{A}_{f\text{c}}$
(does not contain zero precoder), which avoids the trivial case of
non-JT. The product sets of the action spaces are $\mathcal{A}_{\text{uc}}\triangleq\mathcal{A}_{m\text{uc}}\times\mathcal{A}_{f\text{uc}}$
and $\mathcal{A}_{\text{c}}\triangleq\mathcal{A}_{m\text{c}}\times\mathcal{A}_{f\text{c}}$.
The utility functions of the two players in the uncoordinated case
(resp. coordinated case) are $R_{\text{uc}}^{m}$ and $R_{\text{uc}}^{f}$
(resp. $R_{\text{c}}^{m}$ and $R_{\text{c}}^{f}$). The joint action
of $\mathtt{G}_{1}$ is $\mathbf{V}\triangleq\left(\mathbf{V}_{m},\mathbf{V}_{f}\right)\in\mathcal{A}_{\text{uc}}$
where $\mathbf{V}_{m}\in\mathcal{A}_{m\text{uc}}$ and $\mathbf{V}_{f}\in\mathcal{A}_{f\text{uc}}$
and the joint action of $\mathtt{G}_{2}$ is $\mathbf{V}\in\mathcal{A}_{\text{c}}$
such that $\mathbf{V}\triangleq\left(\mathbf{V}_{m},\mathbf{V}_{f}\right)$
where $\mathbf{V}_{m}\in\mathcal{A}_{m\text{c}}$ and $\mathbf{V}_{f}\in\mathcal{A}_{f\text{c}}$.
Note that FBS's action $\mathbf{V}_{f}\in\mathcal{A}_{f\text{c}}$,
consists of two precoders $\mathbf{V}_{f}\triangleq\left(\mathbf{V}_{if},\mathbf{V}_{jf}\right)$.
The MBS (resp. FBS) has identical maximum transmit power in both uncoordinated
and coordinated cases, i.e., for FBS,{\footnotesize{ }}$\underset{\mathbf{V}\in\mathcal{A}_{f\text{uc}}}{\text{max }}\left\{ \text{Trace}\left(\mathbf{V}^{H}\mathbf{V}\right)\right\} =$$\underset{\mathbf{V}_{f}\in\mathcal{A}_{f\text{c}}}{\text{max }}\left\{ \text{Trace}\left(\mathbf{V}_{if}\mathbf{V}_{if}^{H}\right)+\right.\allowbreak\left.\text{Trace}\left(\mathbf{V}_{jf}\mathbf{V}_{jf}^{H}\right)\right\} $
and analogously for MBS.

Now one possesses all the ingredients necessary to define the non-cooperative
games, $\mathtt{G}_{1}$ and $\mathtt{G}_{2}$. The uncoordinated
game is give by the tuple $\mathtt{G}_{1}\triangleq\left\langle \mathcal{N},\mathcal{A}_{\text{uc}},\left\{ R_{\text{uc}}^{m},R_{\text{uc}}^{f}\right\} \right\rangle $.
The game when the two systems are in coordination is $\mathtt{G}_{2}\triangleq\left\langle \mathcal{N},\mathcal{A}_{\text{c}},\left\{ R_{\text{c}}^{m},R_{\text{c}}^{f}\right\} \right\rangle $. 

Let us set aside the above defined two games for a moment, we come
back to them shortly. To analyze the coordinated system one must utilize
coalitional games from the cooperative game theory. The most widely
used solution concept in coalitional games is the \emph{core.} In
order for the two BSs to coordinate the core of the coalition game
must be nonempty. A nonempty core implies that the grand coalition,
which includes all the players, has a value, which is divisible among
the players so that no other partition of subsets of players can give
a better value to any of the players. The analysis of the core requires
that the cooperative game has TU, which means that the sum utility
of the coalition (the two cells in this case) renders itself to be
shared between the members. But one observes, from the system model,
that the sum rate of the coordinated system is not arbitrarily transferable
between the two players. Therefore we follow a usual trick employed
in such situations, introduce a monitory transfer i.e., payment, between
the macro and femto systems. It is imperative to understand that such
a monitory transfer is not merely a tool to make the problem amenable
to coalitional game analysis, but also has an important engineering
and economic aspect: coordination between the systems require sharing
power with external users and communication of symbol information
and channel state information (CSI) between the BSs. Such transactions
have to be compensated in any practical system in order to provide
an incentive to take part in CoMP. After introducing the payment $c$,
the utility of MBS, $U_{\text{c}}^{m}$, and FBS, $U_{\text{c}}^{f}$,
is given by (\ref{eq:MBSutil}). The payment is of units of rate,
which can be interpreted in monitory terms as applicable. 
\begin{align}
U_{\text{c}}^{m}\triangleq & R_{\text{c}}^{m}-c,\qquad U_{\text{c}}^{f}\triangleq R_{\text{c}}^{f}+c.\label{eq:MBSutil}
\end{align}

A coalitional game in characteristic form requires a set of players
and a value function \cite{Nisan2007}. In this paper the set of players
is $\mathcal{\mathcal{N}}$, which has three nonempty subsets. 

To define the value function we revisit the games $\mathtt{G}_{1}$
and $\mathtt{G}_{2}$. There are multiple definitions of equilibria
for non-cooperative games. This research is interested in\emph{ }CE,
which is a generalization of\emph{ }NE \cite{Nisan2007}.
\begin{defn}
CE of the game $\mathtt{G}_{1}$ is a probability distribution $\tilde{p}_{\text{uc}}\left(\cdot\right)$
on the joint action space $\mathcal{A}_{\text{uc}}$ such that $\forall$
$\mathbf{V}\in\mathcal{A}_{\text{uc}}$, $\forall$ $\mathbf{V}_{m}^{\prime}\in\mathcal{A}_{m\text{uc}}$,
and $\forall$ $\mathbf{V}_{f}^{\prime}\in\mathcal{A}_{f\text{uc}}$
\begin{align}
\sum_{\mathbf{V}:\mathbf{V}_{f}\in\mathcal{A}_{f\text{uc}}}\tilde{p}_{\text{uc}}\left(\mathbf{V}\right)R_{\text{uc}}^{m}\left(\mathbf{V}\right)\geq & \sum_{\mathbf{V}:\mathbf{V}_{f}\in\mathcal{A}_{f\text{uc}}}\tilde{p}_{\text{uc}}\left(\mathbf{V}\right)R_{\text{uc}}^{m}\left(\mathbf{V}_{m}^{\prime},\mathbf{V}_{f}\right),\label{eq:CEg1m}\\
\sum_{\mathbf{V}:\mathbf{V}_{m}\in\mathcal{A}_{m\text{uc}}}\tilde{p}_{\text{uc}}\left(\mathbf{V}\right)R_{\text{uc}}^{f}\left(\mathbf{V}\right)\geq & \sum_{\mathbf{V}:\mathbf{V}_{m}\in\mathcal{A}_{m\text{uc}}}\tilde{p}_{\text{uc}}\left(\mathbf{V}\right)R_{\text{uc}}^{f}\left(\mathbf{V}_{f}^{\prime},\mathbf{V}_{m}\right).\label{eq:CEg1f}
\end{align}
Similarly we define the CE of the game $\mathtt{G}_{2}$, the probability
distribution $\tilde{p}_{\text{c}}\left(\cdot\right)$ on the action
space $\mathcal{A}_{\text{c}}$, which satisfies $\forall$ $\mathbf{V}\in\mathcal{A}_{\text{c}}$,
$\forall\,\mathbf{V}_{m}^{\prime}\in\mathcal{A}_{m\text{c}}$, and
$\forall\,\mathbf{V}_{f}^{\prime}\in\mathcal{A}_{f\text{c}}$ 
\begin{align}
\sum_{\mathbf{V}:\mathbf{V}_{f}\in\mathcal{A}_{f\text{c}}}\tilde{p}_{\text{c}}\left(\mathbf{V}\right)R_{\text{c}}^{m}\left(\mathbf{V}\right) & \geq\sum_{\mathbf{V}:\mathbf{V}_{f}\in\mathcal{A}_{f\text{c}}}\tilde{p}_{\text{c}}\left(\mathbf{V}\right)R_{\text{c}}^{m}\left(\mathbf{V}_{m}^{\prime},\mathbf{V}_{f}\right),\label{eq:CEg2m}\\
\sum_{\mathbf{V}:\mathbf{V}_{m}\in\mathcal{A}_{m\text{c}}}\tilde{p}_{\text{c}}\left(\mathbf{V}\right)R_{\text{c}}^{f}\left(\mathbf{V}\right) & \geq\sum_{\mathbf{V}:\mathbf{V}_{m}\in\mathcal{A}_{m\text{c}}}\tilde{p}_{\text{c}}\left(\mathbf{V}\right)R_{\text{c}}^{f}\left(\mathbf{V}_{f}^{\prime},\mathbf{V}_{m}\right).\label{eq:CEg2f}
\end{align}

\end{defn}
While a finite game is guaranteed to have at least one CE, in most
cases there are an infinite set of CE \cite{Nisan2007}. Out of this
set of CE this paper choose the equilibrium, which maximizes the expected
sum rate. The \emph{maximum expected sum rate correlated equilibrium}
(MESR-CE) of game $\mathtt{G}_{1}$ is the probability distribution
obtained through solving the following linear system;

\begin{align}
\underset{\mathbf{p}_{\text{uc}}}{\text{maximize }} & \sum_{\mathbf{V}\in\mathcal{\mathcal{A}_{\text{uc}}}}p_{\text{uc}}\left(\mathbf{V}\right)\left(R_{\text{uc}}^{m}\left(\mathbf{V}\right)+R_{\text{uc}}^{f}\left(\mathbf{V}\right)\right),\nonumber \\
\text{subject to} & \left(\ref{eq:CEg1m}\right),\,\left(\ref{eq:CEg1f}\right),\label{eq:lp1}\\
 & \sum_{\mathbf{\mathbf{V}}\in\mathcal{\mathcal{A}_{\text{uc}}}}p_{\text{uc}}\left(\mathbf{V}\right)=1,\nonumber \\
 & p_{\text{uc}}\left(\mathbf{V}\right)\geq0,\,\forall\,\mathbf{V}\in\mathcal{\mathcal{A}_{\text{uc}}},\nonumber 
\end{align}
where $p_{\text{uc}}\left(\mathbf{V}\right)$ is the probability of
joint action $\mathbf{V}\in\mathcal{\mathcal{A}_{\text{uc}}}$ and
$\mathbf{p}_{\text{uc}}\triangleq\left(p_{\text{uc}}\left(\mathbf{V}\right)\right)_{\mathbf{v}\in\mathcal{\mathcal{A}_{\text{uc}}}}$.
The expected rate of each player at CE of $\mathtt{G}_{1}$ is

\begin{eqnarray}
R_{\text{uc,cor}}^{m} & \triangleq & \sum_{\mathbf{V}\in\mathcal{\mathcal{A}_{\text{uc}}}}\tilde{p}_{\text{uc}}\left(\mathbf{V}\right)R_{\text{uc}}^{m}\left(\mathbf{V}\right),\label{eq:exrateUCm}\\
R_{\text{uc,cor}}^{f} & \triangleq & \sum_{\mathbf{V}\in\mathcal{\mathcal{A}_{\text{uc}}}}\tilde{p}_{\text{uc}}\left(\mathbf{V}\right)R_{\text{uc}}^{f}\left(\mathbf{V}\right),\label{eq:exrateUCf}
\end{eqnarray}
where $\tilde{p}_{\text{uc}}\left(\cdot\right)$ is the MESR-CE solution
of the linear program (\ref{eq:lp1}).

Analogously one can obtain the MESR-CE of game $\mathtt{G}_{2}$ as
the solution to the following linear system; 

\begin{align}
R_{\text{c, cor}}\triangleq\underset{\mathbf{p}_{\text{c}}}{\text{maximize }} & \sum_{\mathbf{V}\in\mathcal{A}_{\text{c}}}p_{\text{c}}\left(\mathbf{V}\right)\left(R_{\text{c}}^{m}\left(\mathbf{V}\right)+R_{\text{c}}^{f}\left(\mathbf{V}\right)\right),\nonumber \\
\text{subject to} & \left(\ref{eq:CEg2m}\right),\,\left(\ref{eq:CEg2f}\right),\label{eq:lp2}\\
 & \sum_{\mathbf{V}\in\mathcal{A}_{\text{c}}}p_{\text{c}}\left(\mathbf{V}\right)=1,\nonumber \\
 & p_{\text{c}}\left(\mathbf{V}\right)\geq0,\,\forall\,\mathbf{V}\in\mathcal{A}_{\text{c}},\nonumber 
\end{align}
where $\mathbf{p}_{\text{c}}\triangleq\left(p_{\text{c}}\left(\mathbf{V}\right)\right)_{\mathbf{V}\in\mathcal{\mathcal{A}_{\text{c}}}}$.
Let $\tilde{p}_{\text{c}}\left(\cdot\right)$ be the MESR-CE distribution
of game $\mathtt{G}_{2}$. The expected rate of each player at CE
of $\mathtt{G}_{2}$ is

\begin{eqnarray}
R_{\text{c,cor}}^{m} & \triangleq & \sum_{\mathbf{V}\in\mathcal{\mathcal{A}_{\text{c}}}}\tilde{p}_{\text{c}}\left(\mathbf{V}\right)R_{\text{c}}^{m}\left(\mathbf{V}\right),\label{eq:exrateCm}\\
R_{\text{c,cor}}^{f} & \triangleq & \sum_{\mathbf{V}\in\mathcal{\mathcal{A}_{\text{c}}}}\tilde{p}_{\text{c}}\left(\mathbf{V}\right)R_{\text{c}}^{f}\left(\mathbf{V}\right).\label{eq:exrateCf}
\end{eqnarray}

Now the value function $v\left(\cdot\right)$ of the coalition game
is as follows;
\begin{equation}
v\left(\mathcal{S}\right)\triangleq\begin{cases}
R_{\text{uc,cor}}^{m} & \mathcal{S}=\left\{ \text{MBS}\right\} ,\\
R_{\text{uc, cor}}^{f} & \mathcal{S}=\left\{ \text{FBS}\right\} ,\\
R_{\text{c, cor}} & \mathcal{S}=\mathcal{N}.
\end{cases}\label{eq:value}
\end{equation}

At this point let us recap the development of this section so far:
in the above definition of the value function $v\left(\mathcal{S}\right)$,
$R_{\text{uc,cor}}^{m}$ in (\ref{eq:exrateUCm}) (resp. $R_{\text{uc, cor}}^{f}$
in (\ref{eq:exrateUCf})) is the expected rate obtained by the macro
system (resp. femto system) while playing the MESR-CE in $\mathtt{G}_{1}$.
On the other hand the value of the grand coalition, $R_{\text{c, cor}}$
in (\ref{eq:lp2}), is the MESR of the two BSs while playing the MESR-CE
in $\mathtt{G}_{2}$. Then the coalitional game in characteristic
form is defined by the tuple $\mathtt{G}_{3}\triangleq\left\langle \mathcal{N},v\left(\cdot\right)\right\rangle $. 
\begin{defn}
The core is the set of allocations such that no subgroup within the
coalition can do better by leaving to form other coalitions \cite{Nisan2007}.
\end{defn}
In our game the set of allocations are $U_{\text{c}}^{m}$ and $U_{\text{c}}^{f}$
in (\ref{eq:MBSutil}), such that $U_{\text{c}}^{m}+U_{\text{c}}^{f}=R_{\text{c, cor}}$.

\subsection{Region of Coordination}

As MUE moves closer to FBS, signal level drops and interference level
rises, hence one expects cooperation with FBS to be preferable to
MBS. Since the sum rate can be apportioned between the two systems
through the monitory transfer, one expects to find a $c$, at which
the core is non empty. The region where the core is non empty is called,
the \emph{region of coordination} or identically \emph{CoMP region.
}In a single input single output (SISO) system a signal to interference
plus noise ratio (SINR) based argument easily demonstrates the existence
of a core but the argument for MIMO requires a bit more analysis.
\begin{prop}
\emph{\label{Claim1} $v\left(\mathcal{N}\right)\geq v\left(\text{MBS}\right)+v\left(\text{FBS}\right)$}
if and only if there exists a payment $c$ such that \emph{$U_{\text{c}}^{m}\geq R_{\text{uc,cor}}^{m}$}
and \emph{$U_{\text{c}}^{f}\geq R_{\text{uc,cor}}^{f}$}. \end{prop}
\begin{IEEEproof}
We provide a constructive proof. By (\ref{eq:MBSutil}) and while
$\mathtt{G}_{2}$ system is in CE the utilities are $U_{\text{c}}^{m}=R_{\text{c,cor}}^{m}-c$\emph{
}and $U_{\text{c}}^{f}=R_{\text{c,cor}}^{m}+c$. Let us consider the
LHS of \emph{iff,} which is equivalent to $R_{\text{c, cor}}\geq R_{\text{uc,cor}}^{m}+R_{\text{uc,cor}}^{f}$,
which implies either $R_{\text{c,cor}}^{m}\geq R_{\text{uc,cor}}^{m}$
or $R_{\text{c,cor}}^{f}\geq R_{\text{uc,cor}}^{f}$ or both. Let
us take the case where $R_{\text{c,cor}}^{m}\geq R_{\text{uc,cor}}^{m}$
and $R_{\text{c,cor}}^{f}\leq R_{\text{uc,cor}}^{f}$, all other cases
can be similarly proven. Then there exists a positive constant $c$
such that $\left(R_{\text{c,cor}}^{m}-c\right)=U_{\text{c}}^{m}\geq R_{\text{uc,cor}}^{m}$
and $\left(R_{\text{c,cor}}^{f}+c\right)=U_{\text{c}}^{f}\geq R_{\text{uc,cor}}^{f}$
since $\left(R_{\text{c,cor}}^{m}-c\right)+\left(R_{\text{c,cor}}^{f}+c\right)\geq\allowbreak R_{\text{uc,cor}}^{m}+R_{\text{uc,cor}}^{f}$.
Converse (RHS$\implies$LHS) is proven simply by summing the two inequalities
\emph{$U_{\text{c}}^{m}\geq R_{\text{uc,cor}}^{m}$ and $U_{\text{c}}^{f}\geq R_{\text{uc,cor}}^{f}$.}
\end{IEEEproof}

Proposition \ref{Claim1} claims that $R_{\text{c, cor}}\geq R_{\text{uc,cor}}^{m}+R_{\text{uc,cor}}^{m}$
is a necessary and sufficient condition for the core of $\mathtt{G}_{3}$
to be nonempty. 

In order to establish the final result we need the following propositions. 
\begin{prop}
\label{claim2}\emph{$R_{\text{uc}}^{m}$} is monotonically decreasing
in \emph{$d_{im}$} and monotonically increasing in \emph{$d_{if}$}.\end{prop}
\begin{IEEEproof}
The proof depends on Loewner ordering of positive semidefinite (PSD)
matrices (\cite{horn2012matrix} 7.7). For two PSD matrices $\mathbf{A}$,
$\mathbf{B}$, we write $\mathbf{A}\succeq\mathbf{B}$ (resp. $\mathbf{A}\succ\mathbf{B}$)
if $\mathbf{A}-\mathbf{B}\succeq0$ is PSD (resp. $\mathbf{A}-\mathbf{B}\succ0$
positive definite (PD)). Let $\mathbf{A}\triangleq\mathbf{H}_{im}\mathbf{V}_{m}\mathbf{V}_{m}^{H}\mathbf{H}_{im}^{H}$,
$\mathbf{B}\triangleq\mathbf{H}_{if}\mathbf{V}_{f}\mathbf{V}_{f}^{H}\mathbf{H}_{if}^{H}$
and $\mathbf{C}\triangleq\sigma^{2}\mathbf{I}_{R}$. $\mathbf{A},\mathbf{B}$
are PSD and $\mathbf{C}$ is PD, also $d_{if}^{-2\alpha}\mathbf{B}+\mathbf{C}$
is PD. Then the capacity of maro-system (\ref{eq:MUEcap}) can be
reformulated as {\small{
\[
R_{\text{uc}}^{m}=\log\det\left(\mathbf{I}_{R}+d_{im}^{-2\alpha}\left(d_{if}^{-2\alpha}\mathbf{B}+\mathbf{C}\right)^{-\frac{1}{2}}\mathbf{A}\left(d_{if}^{-2\alpha}\mathbf{B}+\mathbf{C}\right)^{\frac{-1}{2}}\right).
\]
}}Let $0<d_{im}<d_{im}^{\prime}$, so (\ref{eq:claim2ineq}) \emph{(see
page above)} holds, therefore the determinant of (\ref{eq:detUC})
is no less than the determinant of (\ref{eq:detprimeUC}), which implies
that the determinant is monotonically decreasing in $d_{im}$. 
\begin{equation}
\mathbf{I}_{R}+d_{im}^{\prime-2\alpha}\left(d_{if}^{-2\alpha}\mathbf{B}+\mathbf{C}\right)^{\frac{-1}{2}}\mathbf{A}\left(d_{if}^{-2\alpha}\mathbf{B}+\mathbf{C}\right)^{\frac{-1}{2}}.\label{eq:detprimeUC}
\end{equation}
 
\begin{equation}
\mathbf{I}_{R}+d_{im}^{-2\alpha}\left(d_{if}^{-2\alpha}\mathbf{B}+\mathbf{C}\right)^{\frac{-1}{2}}\mathbf{A}\left(d_{if}^{-2\alpha}\mathbf{B}+\mathbf{C}\right)^{\frac{-1}{2}}.\label{eq:detUC}
\end{equation}
 Next we reformulate (\ref{eq:MUEcap}), 
\[
R_{\text{uc}}^{m}=\log\det\left(\mathbf{I}_{R}+d_{im}^{-2\alpha}\mathbf{A}^{\frac{1}{2}}\left(d_{if}^{-2\alpha}\mathbf{B}+\mathbf{C}\right)^{-1}\mathbf{A}^{\frac{1}{2}}\right),
\]
 and let $0<d_{if}<d_{if}^{\prime}$.{\small{ }}Then (\ref{eq:claim2ineq2})
holds and by a similar argument to above we have that the determinant
is increasing in $d_{if}$. This completes the proof.
\end{IEEEproof}

Now we consider the properties of $R_{\text{c}}^{m}.$
\begin{prop}
\label{claim3}$R_{\text{c}}^{m}\left(\cdot\right)$ is monotonically
increasing in \emph{$d_{if}^{-2\alpha}$ and $d_{im}^{-2\alpha}$
and is bounded from below by }
\[
\gamma^{m}\left(d_{if},\mathbf{V}_{if}\right)\triangleq\log\det\left(\mathbf{I}_{R}+\frac{1}{\sigma^{2}}d_{if}^{-\alpha}\mathbf{H}_{if}\mathbf{V}_{if}\mathbf{V}_{if}^{H}d_{if}^{-\alpha}\mathbf{H}_{if}^{H}\right).
\]
\end{prop}
\begin{IEEEproof}
The proof utilizes Weyl's inequality for Hermitian matrices \cite{Franklin2000}.
Let us first consider \emph{$d_{if}^{-2\alpha}$}.Suppose $\mathbf{X},$
$\mathbf{Y}$ are two Hermitian matrices of size $n\times n$, then
the Weyl's inequality states that 
\[
\lambda_{i}(\mathbf{X})+\lambda_{n}(\mathbf{Y})\leq\lambda_{i}(\mathbf{X}+\mathbf{Y})\leq\lambda_{i}(\mathbf{X})+\lambda_{1}(\mathbf{Y}),\quad i=1,\ldots,n,
\]
 where $\lambda_{i}(\mathbf{X})$ is the $i^{\mathrm{th}}$ largest
eigenvalue of $\mathbf{X}$, i.e., largest eigenvalue is $\lambda_{1}\left(\mathbf{X}\right)$
and smallest is $\lambda_{n}\left(\mathbf{X}\right)$. If $\mathbf{X}$,
$\mathbf{Y}$ are positive semidefinite (PSD) note that the inequality
reduces to
\[
0\leq\lambda_{i}(\mathbf{X})\leq\lambda_{i}(\mathbf{X}+\mathbf{Y})\leq\lambda_{i}(\mathbf{X})+\lambda_{1}(\mathbf{Y}),\quad i=1,\ldots,n.
\]
Let $\mathbf{X}=\mathbf{I}_{R}+\frac{d_{im}^{-2\alpha}}{\sigma^{2}}\mathbf{H}_{im}\mathbf{V}_{m}\mathbf{V}_{m}^{H}\mathbf{H}_{im}^{H}+\frac{d_{if}^{-2\alpha}}{\sigma^{2}}\mathbf{H}_{if}\mathbf{V}_{if}\mathbf{V}_{if}^{H}\mathbf{H}_{if}^{H}$
and $\mathbf{Y}=\frac{\delta}{\sigma^{2}}\mathbf{H}_{if}\mathbf{V}_{if}\mathbf{V}_{if}^{H}\mathbf{H}_{if}^{H}$
where $\delta\in\mathbb{R}_{+}$. Then monotonicity in \emph{$d_{if}^{-2\alpha}$
}follows from 
\[
0<\det\left(\mathbf{X}\right)=\prod_{i}\lambda_{i}(\mathbf{X})\leq\det\left(\mathbf{X}+\mathbf{Y}\right)=\prod_{i}\lambda_{i}(\mathbf{X}+\mathbf{Y}).
\]
 Similarly the proof extends to \emph{$d_{im}^{-2\alpha}$}. Then
setting $d_{im}^{-2\alpha}=0$ the lower bound is achieved. \end{IEEEproof}
\begin{thm}
\label{lemma1}For some $d>0$ under Assumption 1, $\exists$ \emph{$d_{\text{th}}$}
such that for \emph{$d_{if}\leq d_{\text{th}}$,} the region of cooperation
is nonempty%
\footnote{Distances are absolute values. See Fig. \ref{fig:System-model}, Section
\ref{sec:System-Model} for distance notation.%
}.\end{thm}
\begin{IEEEproof}
Let $\mathbf{V}=\left(\mathbf{V}_{if},\mathbf{V}_{jf}\right)\in\mathcal{A}_{\text{c}}$,
$\mathbf{V}^{\prime}\in\mathcal{A}_{\text{uc}}$ be any two actions
from the respective spaces and let the location of the FUE be fixed
relative to the FBS at $\left(\bar{d}_{jf},\bar{\theta}_{j}\right)$.
Then $R_{\text{c}}^{f}\left(\mathbf{V}\right)$ and $R_{\text{uc}}^{f}\left(\mathbf{V}^{\prime}\right)$
are constants irrespective of location of MUE. Now consider that MUE
moves along a trajectory with decreasing $d_{if}$ and increasing
$d_{im}$. By Proposition \ref{claim2}, $R_{\text{uc}}^{m}\left(\mathbf{V}^{\prime}\right)$
is decreasing. As $d_{if}\rightarrow0$, by Proposition \ref{claim3},
$R_{c}^{m}\left(d_{if},\cdot\right)\rightarrow\infty$. Therefore
there must exist $d_{if}\leq d_{\text{th}}$, such that $R_{c}^{m}\left(d_{if},\cdot\right)+R_{\text{c}}^{f}\left(\mathbf{V}\right)\geq\allowbreak R_{\text{uc}}^{m}\left(\mathbf{V}^{\prime}\right)+R_{\text{uc}}^{f}\left(\mathbf{V}^{\prime}\right)$.
Since the action choice was arbitrary $\exists$ $d_{\text{th}}$
such that, $\underset{\mathbf{V}\in\mathcal{A}_{\text{c}}}{\text{min }}\left(R_{\text{c}}^{m}\left(\mathbf{V}\right)+R_{\text{c}}^{f}\left(\mathbf{V}\right)\right)\geq\allowbreak\underset{\mathbf{V}\in\mathcal{A}_{\text{uc}}}{\text{max }}\left(R_{\text{uc}}^{m}\left(\mathbf{V}\right)+R_{\text{uc}}^{f}\left(\mathbf{V}\right)\right).$

Therefore $\forall$ probability distributions $\tilde{p}_{\text{uc}}$
and $\tilde{p}_{\text{c}}$, we have $\sum_{\mathbf{V}\in\mathcal{A}_{\text{c}}}\tilde{p}_{\text{c}}\left(\mathbf{V}\right)\left(R_{\text{c}}^{m}\left(\mathbf{V}\right)+R_{\text{c}}^{f}\left(\mathbf{V}\right)\right)\geq\allowbreak\sum_{\mathbf{V}\in\mathcal{A}_{\text{uc}}}\tilde{p}_{\text{uc}}\left(\mathbf{V}\right)\left(R_{\text{uc}}^{m}\left(\mathbf{V}\right)+R_{\text{uc}}^{f}\left(\mathbf{V}\right)\right).$
This completes the proof.
\end{IEEEproof}

Theorem \ref{lemma1} together with Proposition \ref{Claim1} suggests
the existence of a region around the FBS where the core is nonempty.
Thus we establish the rationality of CoMP scheme JT.

\section{Numerical Results\label{sec:Numerical-Results}}

\begin{figure}
\includegraphics[scale=0.37]{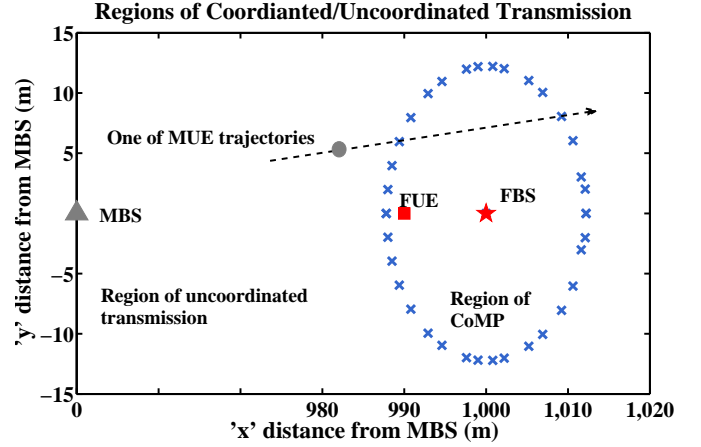}

\caption{Region of coordination.\label{fig:Region-of-coordination}}
\end{figure}

The distances are measured in meters (m), we locate MBS at $\left(0,0\right)$,
FBS at $\left(1000,0\right)$, and FUE $\left(990,0\right)$. Unless
otherwise stated, the default maximum transmit power of MBS is $5$
W and of FBS is $1$ W. The two BSs each has $4$ antennas and each
UE has $2$ antennas. In the coordinated mode of transmission, by
default FBS distributes the power evenly among FUE and MUE. The AWGN
power is set at $10^{-4}$ W. In the Fig. \ref{fig:Region-of-coordination}
MUE moves from far negative $x$ region towards the FBS in linear
trajectories. One such trajectory is shown in the figure. The region
where coordination is preferred over uncoordinated transmission is
marked. The symmetry in the region is due to the use of symmetric
channel matrices on either side of the FBS. 

\begin{figure}
\includegraphics[scale=0.37]{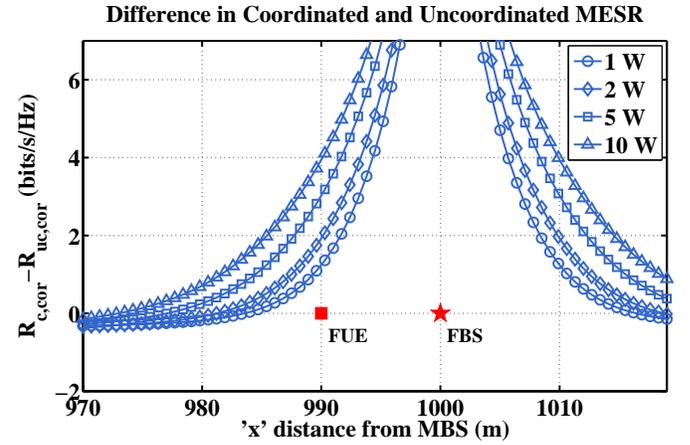}

\caption{Dependence of CoMP region on FBS transmit power.\label{fig:Dependence-of-CoMP}}
\end{figure}

\begin{figure}
\includegraphics[scale=0.37]{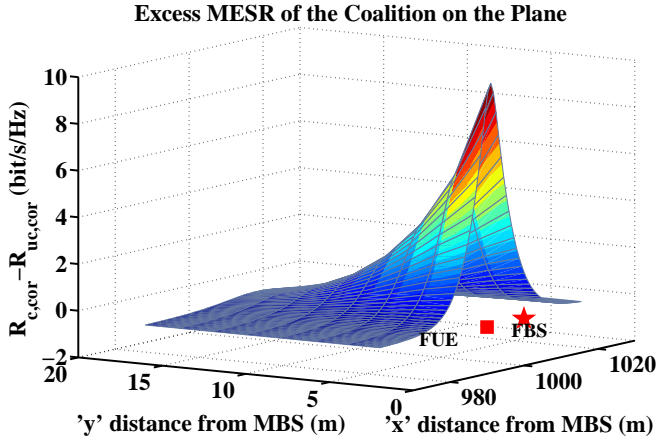}

\caption{Geometric distribution of value of coalition, $v\left(\mathcal{N}\right),$
over uncoordinated system. \label{fig:Geometric-distribution-of}}
\end{figure}

In the rest of the figures the trajectory of the MUE is on the $x$
axis ($y$ coordinate is $0$). Fig. \ref{fig:Dependence-of-CoMP}
denotes the expansion of the CoMP region as the FBS transmit power
increases. One also sees from the figure that $R_{\text{c, cor}}$
far exceeds $R_{\text{uc, cor}}$ as MUE approaches FBS. Fig. \ref{fig:Geometric-distribution-of}
shows, on the plan of $\left(x,\, y\right)$, the excess value of
the coalition over the value of uncoordinated system.

\begin{figure}
\includegraphics[scale=0.37]{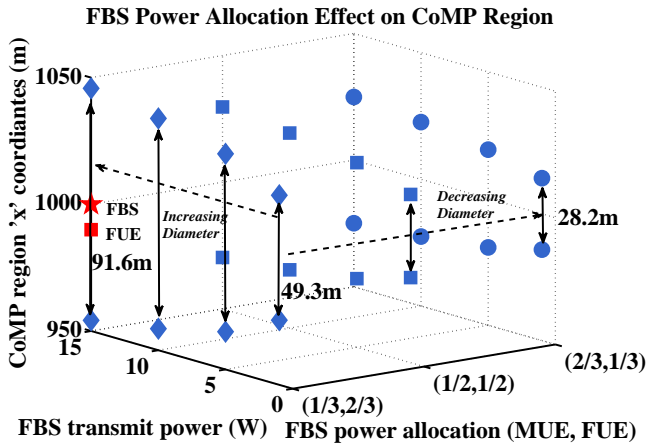}

\caption{The variation of the diameter of the CoMP region with FBS power allocation
ratio $\left(a,b\right)$. Here $\left(a,b\right)$ implies $a$ and
$b$ fractions of power are assigned to MUE and FUE respectively.\label{fig:The-variation-of}}
\end{figure}

Fig. \ref{fig:The-variation-of} demonstrates that as the amount of
power allocated to MUE increases the \emph{diameter} of the coordination
region shrinks. The term diameter is loosely used to mean the distance
between the entry point and exit point of CoMP region when the MUE's
trajectory is on $x$ axis ($y$ coordinate $0$). Consider the two
MUE power ratios of $a$ and $c$ such that $c>a$. Then the explanation
for the phenomenon seen in Fig. \ref{fig:The-variation-of} is that
while operating at ratio $c$ if the FBS switches to CoMP at the coordination
boundary of the ratio $a$ then the reduction of FUE rate is higher
than the increase in MUE rate as still MUE is further away from FBS
than FUE, thus discouraging the formation of the coalition till MUE
moves closer to FBS.

\section{Conclusion\label{sec:Conclusion}}

This paper considered the downlink of a HetNet consisting of a maro-
and a femtocell. Two non-cooperative games were devised. The first
game, $\mathtt{G}_{1}$, had the two cells in competition. In the
second game, $\mathtt{G}_{2}$, the cells were in coordination (CoMP).
In each game the cells operated in the respective maximum expected
sum rate-correlated equilibria (MESR-CE). Then a third game, $\mathtt{G}_{3}$,
was defined which is a coalition game in characteristic form with
transferable utility. In $\mathtt{G}_{3}$ the value of the coalition
was allowed to be arbitrarily transferred between the two cells via
a payment. The solution mechanism of \emph{core,} was used in the
coalitional games $\mathtt{G}_{3}$ with value function based on MESR-CE
of $\mathtt{G}_{1}$ and $\mathtt{G}_{2}$. Then the paper proved
the existence of a region where the core of the game $\mathtt{G}_{3}$
is nonempty, which demonstrates that CoMP is a rational decision in
some region and the CoMP decision making is reduced to identifying
a threshold separation $d_{\text{th}}$. CoMP decision mechanisms
for more complex channel models with more than two cells can be considered
in future work.

\subsubsection*{Acknowledgment}

We thank Prof. N. Rajatheva of University of Oulu, Finland and Dr.
F. Poloni of University of Pisa, Italy, for their valuable insight
in proofs. We also thank the StackExchange forums for providing a
platform of discussion.

\bibliographystyle{IEEEtran}
\bibliography{IEEEabrv,icc13,coaref}
\balance
\end{document}